\documentclass[aps,prd,nofootinbib,onecolumn,preprintnumbers,floatfix,superscriptaddress,bibliography]{revtex4-2}
\usepackage[colorlinks, linkcolor=red, anchorcolor=blue, citecolor=blue, colorlinks=true, urlcolor=blue]{hyperref}
\usepackage{amsmath}
\usepackage{graphicx}
\usepackage{fontawesome}
\usepackage{ulem}
\usepackage{bm}
\usepackage{xspace}
\usepackage{amssymb}
\usepackage{gensymb}
\usepackage{multirow}
\usepackage{threeparttable}
\usepackage{amsthm}
\usepackage{mathrsfs}
\usepackage{indentfirst}
\usepackage{subfigure}
\usepackage[figuresright]{rotating}
\usepackage{epstopdf}
\usepackage{breakurl}
\usepackage{lipsum}
\usepackage{cancel}
\usepackage{slashed}

\begin{document}

\title{Gravitational waves and primordial black holes from axion domain walls\\ in level crossing}

\author{Hai-Jun Li} 
\email{lihaijun@itp.ac.cn}
\affiliation{Key Laboratory of Theoretical Physics, Institute of Theoretical Physics, Chinese Academy of Sciences, Beijing 100190, China}
 
\author{Yu-Feng Zhou}
\email{yfzhou@itp.ac.cn}
\affiliation{Key Laboratory of Theoretical Physics, Institute of Theoretical Physics, Chinese Academy of Sciences, Beijing 100190, China}
\affiliation{School of Physical Sciences, University of Chinese Academy of Sciences, Beijing 100049, China}
\affiliation{School of Fundamental Physics and Mathematical Sciences, Hangzhou Institute for Advanced Study, UCAS, Hangzhou 310024, China}
\affiliation{International Centre for Theoretical Physics Asia-Pacific, Beijing/Hangzhou, China}

\preprint{ITP-24-003}

\date{\today}

\begin{abstract}

In this paper, we investigate the nano-Hertz gravitational waves (GWs) emission and the massive primordial black holes (PBHs) formation from the light QCD axion scenario.
We consider the axion domain walls formation from the level crossing induced by the mass mixing between the light $Z_{\mathcal N}$ QCD axion and axion-like particle.
A general mixing case is considered that the heavy and light mass eigenvalues do not necessarily have to coincide with the axion masses.
In order to form the domain walls, the axions should start to oscillate slightly before the level crossing.
The domain walls must annihilate before dominating the Universe to avoid the cosmological catastrophe.
Then we focus our attention on the GWs emission from the domain walls annihilation and the PBHs formation from the domain walls collapse.
We show the predicted GWs spectra with the peak frequency $\sim 0.2\, \rm nHz$ and the peak amplitude $\sim 5\times 10^{-9}$, which can be tested by the future pulsar timing array projects.
In addition, during the domain walls annihilation, the closed walls could shrink to the Schwarzschild radius and collapse into the PBHs.
We find that the PBHs in the mass range of $\mathcal{O}(10^5-10^8) M_\odot$ could potentially form in this scenario and account for a small fraction $\sim 10^{-5}$ of the cold dark matter.


\end{abstract}
\maketitle

 
\section{Introduction}

Gravitational wave (GW) is one of the most effective observational probes of the early Universe \cite{Maggiore:1999vm, Maggiore:2007ulw}.
The direct observations of the GWs by LIGO \cite{LIGOScientific:2016aoc, LIGOScientific:2016sjg, LIGOScientific:2017bnn} have brought tremendous development to astrophysics and cosmology.
Very recently, the pulsar timing array (PTA) projects (NANOGrav \cite{NANOGrav:2023gor, NANOGrav:2023hfp, NANOGrav:2023hvm}, EPTA \cite{EPTA:2023sfo, EPTA:2023akd, EPTA:2023fyk}, PPTA \cite{Reardon:2023gzh, Reardon:2023zen}, and CPTA \cite{Xu:2023wog}) released their latest data, which provides strong evidence for the presence of nano-Hertz stochastic GWs.
There are various cosmological sources of the GWs, such as the primordial amplification of the vacuum fluctuations \cite{Grishchuk:1974ny, Starobinsky:1979ty, Smith:2005mm}, the cosmological phase transitions \cite{Witten:1984rs, Kamionkowski:1993fg}, the cosmic strings \cite{Vilenkin:1981bx, Accetta:1988bg, Caldwell:1991jj}, the domain walls \cite{Vilenkin:1981zs, Preskill:1991kd, Chang:1998tb, Gleiser:1998na}, and the preheating after inflation \cite{Khlebnikov:1997di, Easther:2006gt, Garcia-Bellido:2007fiu, Dufaux:2008dn}, etc.
See $\rm e.g.$ refs.~\cite{Blanchet:2013haa, Sasaki:2018dmp, Caprini:2018mtu} for recent reviews of the GWs.

After the observations of the GWs, the primordial black holes (PBHs) have recently gathered significant interest, which are attractive cold dark matter (DM) candidates.  
There are several scenarios for the PBHs formation in the early Universe, such as the large density fluctuations produced during inflation \cite{Carr:1975qj, Ivanov:1994pa, Garcia-Bellido:1996mdl, Alabidi:2009bk}, the cosmological phase transitions \cite{Kodama:1982sf, Hawking:1982ga, Baker:2021nyl, Gouttenoire:2023naa}, and the collapse of the false vacuum bubbles \cite{Deng:2017uwc, Deng:2020mds, Maeso:2021xvl}, the cosmic strings \cite{Hawking:1987bn, Polnarev:1988dh, MacGibbon:1997pu}, the domain walls \cite{Rubin:2001yw, Garriga:2015fdk, Deng:2016vzb, Gouttenoire:2023gbn}, and the post-inflationary scalar field fragmentations \cite{Cotner:2016cvr, Cotner:2018vug, Cotner:2019ykd}, etc.
See also $\rm e.g.$ refs.~\cite{Carr:2020gox, Carr:2020xqk, Green:2020jor} for recent reviews of the PBHs.

Here we consider the domain wall as a cosmological source of the GWs and PBHs.
The domain walls are two-dimensional topological defects that may arise in the early Universe involving the spontaneous breakdown of a discrete symmetry \cite{Kibble:1976sj, Sikivie:1982qv}.
For the long-lived domain walls, their evolution with the cosmic expansion is slower than that of radiation or matter, and they will eventually dominate the total energy density of the Universe, which conflicts with the standard cosmology \cite{Zeldovich:1974uw}.
This is the so-called domain wall problem.
To avoid such a cosmological catastrophe, one can make the domain walls disappear or prevent their formation before dominating the Universe \cite{Lazarides:1982tw, Gelmini:1988sf, Larsson:1996sp}.

The GWs emitted by the domain walls annihilation can be characterized by their peak frequency and peak amplitude, in which the former one is determined by the domain walls annihilation time, and the latter one is determined by the energy density of the domain walls \cite{Hiramatsu:2010yz, Kawasaki:2011vv, Hiramatsu:2013qaa}. 
Here we focus our attention on the GWs emission from the domain walls annihilation, especially the axion domain walls \cite{Higaki:2016jjh, Caputo:2019wsd, Chiang:2020aui, Gelmini:2021yzu, Sakharov:2021dim, Ferreira:2022zzo, Blasi:2022ayo, Kanno:2023kdi, Kitajima:2023cek, Blasi:2023sej, Gouttenoire:2023ftk, Lu:2023mcz, Gelmini:2023kvo, Ge:2023rce, Li:2023gil, Kitajima:2023kzu, Chen:2021wcf, Gelmini:2022nim, Gelmini:2023ngs}.
In addition, during the annihilation, the closed domain walls could shrink to the Schwarzschild radius and then collapse into the PBHs.
In this case, the PBHs will form when this Schwarzschild radius is comparable to the cosmic time.
See also $\rm e.g.$ refs.~\cite{Ferrer:2018uiu, Ge:2019ihf, Kitajima:2020kig, Chen:2021wcf, Choi:2022btl, Li:2023det, Li:2023zyc, Gelmini:2022nim, Gelmini:2023ngs, Kasai:2023ofh, Ge:2023rrq, Dunsky:2024zdo} for recent PBH mechanisms with the framework of the QCD axion or axion-like particle (ALP).

\begin{figure*}[t]
\centering
\includegraphics[width=0.78\textwidth]{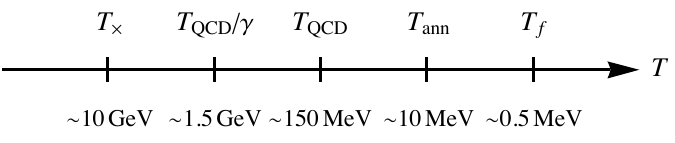}
\caption{The main cosmic temperatures related to this work, where $T_\times$ represents the level crossing temperature, $T_{\rm QCD}$ represents the QCD phase transition critical temperature, $T_{\rm ann}$ represents the domain walls annihilation temperature, $T_f$ represents the PBH formation temperature, and $\gamma\in(0,1)$ is a temperature parameter.
The temperature is decreasing from left to right.
Note that the ticks are shown just for illustrative purposes.
See also the text for more details.}
\label{fig_Tall}
\end{figure*}

In this paper, we investigate the nano-Hertz GWs emission and the massive PBHs formation from the light QCD axion scenario discussed in ref.~\cite{Li:2023uvt}.
The basic idea is considering the axion domain walls formation from the level crossing induced by the light $Z_{\mathcal N}$ QCD axion.
In the $Z_{\mathcal N}$ axion scenario \cite{Hook:2018jle}, the $\mathcal N$ mirror worlds are nonlinearly realized by the axion field under a $Z_{\mathcal N}$ symmetry, one of which is the Standard Model (SM) world.
The $Z_{\mathcal N}$ axion with the reduced-mass can both solve the strong CP problem with $\mathcal N\geqslant3$ \cite{DiLuzio:2021pxd}, and account for the DM through the trapped+kinetic misalignment mechanism \cite{DiLuzio:2021gos, Co:2019jts}.
Considering the interaction between the $Z_{\mathcal N}$ axion and ALP, the cosmological evolution called the single/double level crossings will occur if there is a non-zero mass mixing between them \cite{Li:2023uvt, Li:2023xkn}. 
In this work we will consider a more general case in the mixing, the heavy and light mass eigenvalues do not necessarily have to coincide with the axion masses, and there is a hierarchy between the two axion decay constants.
In order to form the domain walls in this scenario, the axions should start to oscillate slightly before the level crossing, and the initial oscillation energy density should be large to climb over the barrier of potential \cite{Daido:2015bva}.

To avoid the unacceptable cosmological catastrophe, the domain walls must annihilate before dominating the Universe.
Then we investigate the GWs emitted by the domain walls annihilation and show the predicted GWs spectra, determined by their peak frequency and peak amplitude.
For model parameters with the benchmark values, we have the predicted GWs spectrum with the peak frequency $f_{\rm peak}\sim 0.2\, \rm nHz$ and the peak amplitude $\Omega_{\rm GW}h^2\sim 5\times 10^{-9}$, which can be observable in the future GWs detectors, such as the PTA projects.
Finally, we investigate the PBHs formation from the domain walls collapse.
We consider the domain walls collapse in an approximately spherically symmetric way.
The PBHs will form when the ratio of the Schwarzschild radius to the cosmic time is close to 1, leading to the $\mathcal{O}(10^5-10^8) M_\odot$ (in the solar mass $M_\odot$) massive PBHs as a small fraction $f_{\rm PBH}\sim 10^{-5}$ of the cold DM.
In this regard, we find that these PBHs may also account for the seeds of supermassive black holes (SMBHs) at the high redshift.
See figure~\ref{fig_Tall} for the main cosmic temperatures related to this work.

The rest of this paper is structured as follows.
In section~\ref{sec_axion_domain_walls_light_QCD_axion}, we introduce the light QCD axion scenario and discuss the domain walls formation and annihilation.
In section~\ref{sec_Cosmological_implications}, we investigate the GWs emitted by the domain walls annihilation and the PBHs formation from the domain walls collapse.
Finally, the conclusion is given in section~\ref{sec_conclusion}.

\section{Domain walls form light QCD axion}
\label{sec_axion_domain_walls_light_QCD_axion} 

In this section, we first introduce the light QCD axion and the resulting axion level crossing, then we discuss the domain walls formation and annihilation. 

\subsection{Light QCD axion}

\begin{figure}[t]
\centering
\includegraphics[width=0.65\textwidth]{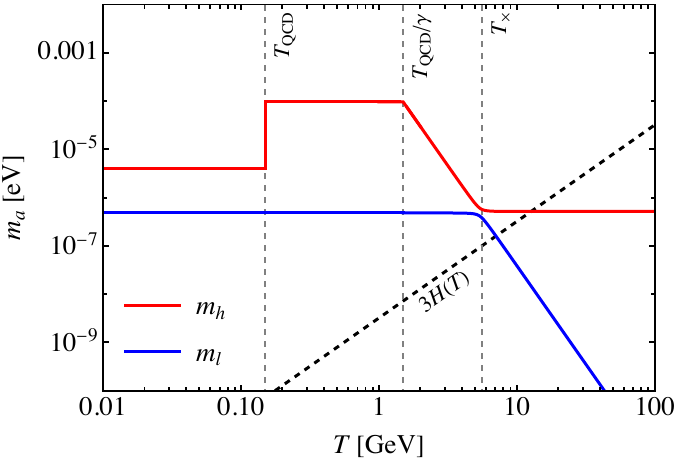}
\caption{The temperature-dependent mass eigenvalues $m_{h,l}$ as functions of the cosmic temperature $T$.
The red and blue solid lines represent $m_h$ and $m_l$, respectively.
The black dashed line represents the Hubble parameter $H(T)$.}
\label{fig_lc}
\end{figure}

In the light $Z_{\mathcal N}$ QCD axion scenario \cite{Hook:2018jle}, the $\mathcal N$ mirror and degenerate worlds that are nonlinearly realized by the axion field under a $Z_{\mathcal N}$ symmetry can coexist with the same coupling strengths as in the SM
\begin{eqnarray}
\mathcal L = \sum_{k=0}^{\mathcal N -1}\left[\mathcal L_{{\rm SM}_k} +\dfrac{\alpha_s}{8\pi}\left(\dfrac{\phi}{f_a}+\dfrac{2\pi k}{\mathcal N}\right)G_k \widetilde{G}_k\right]+ \cdots \, ,
\end{eqnarray}
where $\mathcal L_{{\rm SM}_k}$ represents the copies of the SM total Lagrangian excluding the topological term $G_k \widetilde{G}_k$, $\alpha_s$ is the strong fine structure constant, $\phi$ and $f_a$ are the $Z_{\mathcal N}$ axion field and decay constant, respectively.
In the large $\mathcal N$ limit, the temperature-dependent $Z_{\mathcal N}$ axion mass is given by \cite{DiLuzio:2021pxd, DiLuzio:2021gos}
\begin{eqnarray}
m_a(T)\simeq 
\begin{cases}
m_{a,0}\, , &T\leq T_{\rm QCD}\\
m_{a,\pi}\, , &T_{\rm QCD} < T \leq T_{\rm QCD}/\gamma\\
m_{a,\pi}\left(\dfrac{\gamma T}{T_{\rm QCD}}\right)^{-b}\, . &T>T_{\rm QCD}/\gamma
\end{cases} 
\end{eqnarray}
The zero-temperature $Z_{\mathcal N}$ axion mass $m_{a,0}$ and the defined mass $m_{a,\pi}$ are
\begin{eqnarray}
m_{a,0}&\simeq&\dfrac{m_\pi f_\pi}{f_a}\dfrac{1}{\sqrt[4]{\pi}}\sqrt[4]{\dfrac{1-z}{1+z}}{\mathcal N}^{3/4}z^{\mathcal N/2}\, ,\\
m_{a,\pi}&=&\dfrac{m_\pi f_\pi}{f_a}\sqrt{\dfrac{z}{1-z^2}}\, ,
\end{eqnarray}
where $\gamma\in(0,1)$ is a temperature parameter, $T_{\rm QCD}\simeq150\, \rm MeV$ is the QCD phase transition critical temperature, $b\simeq4.08$ is an index, $m_\pi$ and $f_\pi$ are the mass and decay constant of the pion, respectively, and $z\equiv m_u/m_d\simeq0.48$ is the ratio of up ($m_u$) to down ($m_d$) quark masses. 
The level crossing can take place at the temperature $T_\times$ in the interaction between the $Z_{\mathcal N}$ axion and ALP ($\psi$) with the potential \cite{Li:2023uvt}
\begin{eqnarray}
V(\psi,\phi)=m_A^2 f_A^2\left[1-\cos\left(n\frac{\psi}{f_A}+{\mathcal N}\frac{\phi}{f_a}\right)\right]+\frac{m_a^2(T) f_a^2}{{\mathcal N}^2}\left[1-\cos\left({\mathcal N}\frac{\phi}{f_a}\right)\right]\, ,
\label{V1}
\end{eqnarray}
with the overall scales 
\begin{eqnarray}
\Lambda_1=\sqrt{m_A f_A}\, , \quad \Lambda_2=\sqrt{m_a(T) f_a/{\mathcal N}}\, , 
\end{eqnarray}
where $m_A$ and $f_A$ are the ALP mass and decay constant, respectively, $n$ is a positive integer, $n$ and $\mathcal N$ are also the domain wall numbers.
Through the mass mixing matrix derived from eq.~(\ref{V1}), we can obtain the heavy ($h$) and light ($l$) mass eigenvalues $m_{h,l}$, which are temperature-dependent.
See figure~\ref{fig_lc} for an illustration of the level crossing, and in the following we will consider this case.

Note however that in this work the heavy and light mass eigenvalues do not necessarily have to coincide with the axion masses, which is somewhat different from the case in ref.~\cite{Li:2023uvt}.
In general, there is a hierarchy between the two axion decay constants \cite{Kim:2004rp, Ben-Dayan:2014zsa}. 
Considering two physical fields $\mu$ and $\xi$, in which the first one is the linear combination in the first term of eq.~(\ref{V1}) and the second one is the orthogonal combination, with 
\begin{eqnarray}
\mu&=&\dfrac{f_A f_a}{\sqrt{{\mathcal N}^2 f_A^2 + n^2 f_a^2}}\left(n\dfrac{\psi}{f_A}+{\mathcal N}\dfrac{\phi}{f_a}\right)\, ,\\
\xi&=&\dfrac{f_A f_a}{\sqrt{{\mathcal N}^2 f_A^2 + n^2 f_a^2}}\left(-{\mathcal N}\dfrac{\psi}{f_a}+ n\dfrac{\phi}{f_A}\right)\, ,
\end{eqnarray} 
then the potential in terms of the physical fields can be written as
\begin{eqnarray}
\begin{aligned}
V(\mu,\xi)&=\Lambda_1^4\left[1-\cos\left(\dfrac{\sqrt{{\mathcal N}^2 f_A^2 + n^2 f_a^2}}{f_A f_a}\mu\right)\right]\\
&+\Lambda_2^4\left[1-\cos\left(\dfrac{{\mathcal N}^2 f_A}{f_a \sqrt{{\mathcal N}^2 f_A^2 + n^2 f_a^2}}\mu +\dfrac{n {\mathcal N}}{\sqrt{{\mathcal N}^2 f_A^2 + n^2 f_a^2}}\xi\right)\right]\, ,
\label{V2}
\end{aligned}
\end{eqnarray}
with the effective axion decay constants $f_\mu$ and $f_\xi$
\begin{eqnarray}
f_\mu=\dfrac{f_A f_a}{\sqrt{{\mathcal N}^2 f_A^2 + n^2 f_a^2}}\, ,\quad f_\xi=\dfrac{\sqrt{{\mathcal N}^2 f_A^2 + n^2 f_a^2}}{n {\mathcal N}}\, . 
\end{eqnarray} 

\subsection{Domain walls formation}

In this subsection, we briefly discuss the axion domain walls formation in this scenario.
The domain walls formation from the canonical level crossing case was studied in ref.~\cite{Daido:2015bva}.
It was shown that the formation of domain walls from the level crossing in the axiverse is a common phenomenon.
The onset of axion oscillations is considered slightly before the level crossing temperature $T'_\times$.
In the case the initial axion oscillation energy density is sufficiently large to climb over the barrier of potential, the axion dynamics has therefore a chaotic run-away behavior (also called the axion roulette), which is considered to be accompanied by the domain walls formation. 

In our scenario, we have a similar consideration that the axions start to oscillate slightly before the level crossing
\begin{eqnarray}
T_1 \gtrsim T_\times \, ,
\end{eqnarray} 
where $T_1$ is the axion oscillation temperature given by $m_{h,l}(T)=3H(T)$ with the Hubble parameter
\begin{eqnarray}
H(T)=\sqrt{\dfrac{\pi^2 g_*(T)}{90}}\dfrac{T^2}{m_{\rm Pl}}\, ,
\end{eqnarray} 
where $g_*$ is the number of effective degrees of freedom of the energy density, and $m_{\rm Pl}\simeq 2.44\times10^{18}\, \rm GeV$ is the reduced Planck mass.
Note that here the level crossing occurs earlier than the canonical case with 
\begin{eqnarray}
T_\times \simeq T'_\times/\gamma \, .
\end{eqnarray} 
Then another condition for axion domain walls formation is that the oscillation energy density in the light mass eigenvalue $m_l$ should be larger than the barrier of potential
\begin{eqnarray}
\rho_{l,1} \sim m_{l,1}^2 f_\xi^2  \gtrsim \Lambda_1^4 \sim m_{h,1}^2 f_\mu^2\, ,
\label{condition}
\end{eqnarray} 
where the subscript ``1" corresponds to $T_1$.
Since no cosmic strings are formed, the domain walls without cosmic strings are stable in a cosmological time scale. 

\subsection{Domain walls annihilation}

Then we discuss the domain walls annihilation.
After formation, we consider the dynamics of domain walls is dominated by the tension force.
In the scaling regime, the evolution of walls can be described by the scaling solution \cite{Press:1989yh, Hindmarsh:1996xv, Garagounis:2002kt, Oliveira:2004he}.
In this case, the energy density of domain walls evolves as
\begin{eqnarray}
\rho_{\rm wall}(t)=\mathcal{A}\frac{\sigma_{\rm wall}}{t}\, ,
\end{eqnarray}
where $\mathcal{A}\simeq 0.8 \pm 0.1$ is a scaling parameter obtained from the numerical simulations, and $\sigma_{\rm wall}$ is the tension of domain walls
\begin{eqnarray}
\sigma_{\rm wall}&= 8 m_{h,1} f_\mu^2\simeq 8\zeta \eta^2 m_{a,\pi} f_a^2\, ,
\end{eqnarray} 
where we have defined the parameters 
\begin{eqnarray}
\zeta\equiv m_{h,1} /m_{a,\pi}\ll1\, ,
\end{eqnarray} 
\begin{eqnarray}
\eta \equiv f_A/f_a\simeq f_\mu/f_a \ll1\, .
\end{eqnarray} 
Due to the slower evolution of domain walls with the cosmic expansion compared to radiation or matter, the walls will eventually dominate the Universe, which conflicts with the standard cosmology.
Therefore, to avoid the domain wall problem, they must annihilate before dominating the Universe.
In general, one can introduce an additional small energy difference --``bias"-- between the different vacua to drive the domain walls towards their annihilation.
While in our scenario, the domain walls will become unstable and annihilate due to the natural bias term
\begin{eqnarray}
V_{\rm bias}=\Lambda_b^4\left[1-\cos\left(n\frac{\psi}{f_A}+{\mathcal N}\frac{\phi}{f_a}+\delta\right)\right]\, ,
\end{eqnarray} 
where we have defined a bias parameter $\Lambda_b$, and $\delta$ is a CP phase, which should not spoil the Peccei-Quinn (PQ) solution to the strong CP problem.
Note that the scale $\Lambda_b$ should be several orders of magnitude smaller than the QCD scale, but a too small $\Lambda_b$ may lead to the long-lived domain walls that overclose the Universe.

\begin{figure}[t]
\centering
\includegraphics[width=0.65\textwidth]{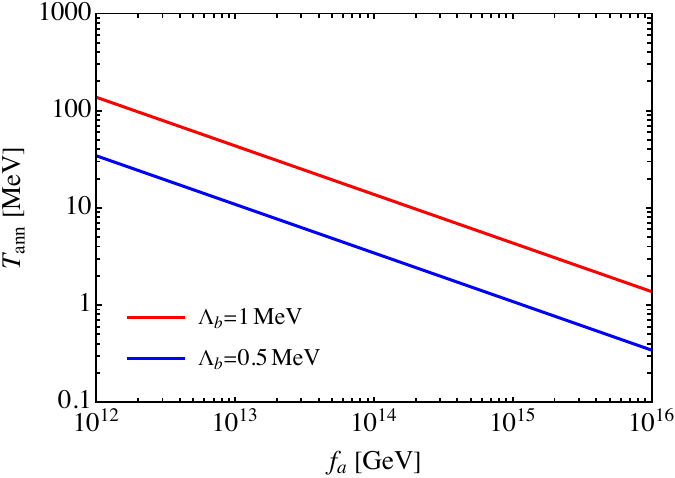}
\caption{The domain walls annihilation temperature $T_{\rm ann}$ as a function of the $Z_{\mathcal N}$ axion decay constant $f_a$.
The red and blue lines represent $\Lambda_b=1\, \rm MeV$ and $0.5\, \rm MeV$, respectively.
Here we take the benchmark values as $\zeta=0.1$ and $\eta=0.1$.}
\label{fig_Tann}
\end{figure} 

\begin{figure}[t]
\centering
\includegraphics[width=0.65\textwidth]{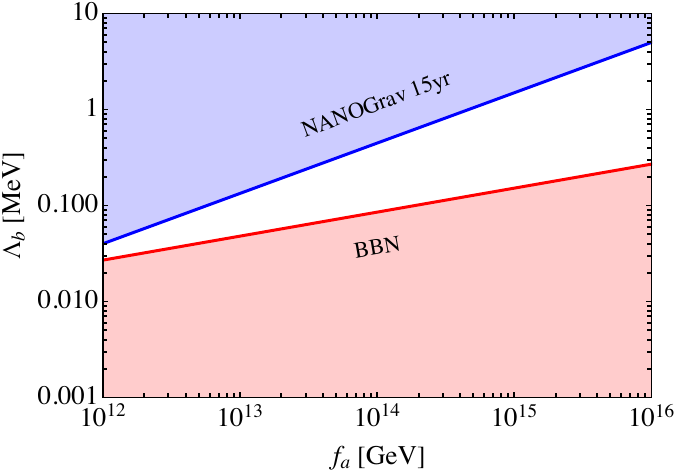}
\caption{The constraints of BBN and NANOGrav 15-year dataset in the $\{f_a, \, \Lambda_b\}$ plane.
The red shadow region represents the constraint set by BBN.
The blue shadow region represents the predicted GWs spectra can not explain the NANOGrav 15-year dataset, as detailed in the text.
Here we set $\zeta=0.1$ and $\eta=0.1$.}
\label{fig_lambda_b}
\end{figure}

Then the resulting volume pressure $p_V\simeq \Lambda_b^4$ accelerates the domain walls towards the higher energy adjacent vacuum, converting the higher energy vacuum into the lower energy vacuum. 
The walls annihilation becomes significant when the pressure produced by the tension $p_T\simeq\rho_{\rm wall}$ is comparable to the volume pressure. 
By taking $p_T\simeq p_V$, we have
\begin{eqnarray}
\mathcal{A}\frac{\sigma_{\rm wall}}{t_{\rm ann}}\simeq \Lambda_b^4\, ,
\end{eqnarray}
where the domain walls annihilation time $t_{\rm ann}$ is given by $H(T_{\rm ann})=1/(2t_{\rm ann})$, corresponding to the domain walls annihilation temperature
\begin{eqnarray}
T_{\rm ann}\simeq 1.4 \, {\rm MeV} \left(\dfrac{\zeta}{0.1}\right)^{-1/2} \left(\dfrac{\eta}{0.1}\right)^{-1} \left(\dfrac{f_a}{10^{16}\, \rm GeV}\right)^{-1/2} \left(\dfrac{\Lambda_b}{1\, \rm MeV}\right)^{2}\, .
\end{eqnarray}
Here we take the benchmark values as $\zeta=0.1$, $\eta=0.1$, $f_a=10^{16}\, \rm GeV$, and $\Lambda_b=1\, \rm MeV$.
In figure~\ref{fig_Tann}, we show the annihilation temperature $T_{\rm ann}$ as a function of the $Z_{\mathcal N}$ axion decay constant $f_a$.
The red and blue lines correspond to the parameter $\Lambda_b=1\, \rm MeV$ and $0.5\, \rm MeV$, respectively.
The other parameters are taken as the benchmark values.
We note that $T_{\rm ann}$ is below the QCD phase transition critical temperature.
Additionally, the domain walls annihilation should before the Big Bang Nucleosynthesis (BBN) epoch to avoid the strong constraints, $\rm i.e.$, $T_{\rm ann}>T_{\rm BBN}\simeq 0.1\, \rm MeV$, then we have
\begin{eqnarray}
\Lambda_b\gtrsim 0.3 \, {\rm MeV} \left(\dfrac{\zeta}{0.1}\right)^{1/4} \left(\dfrac{\eta}{0.1}\right)^{1/2} \left(\dfrac{f_a}{10^{16}\, \rm GeV}\right)^{1/4}\, .
\end{eqnarray} 
We show in figure~\ref{fig_lambda_b} the bias parameter $\Lambda_b$ constrained by BBN in the $\{f_a, \, \Lambda_b\}$ plane with the red shadow region. 
  
\section{Cosmological implications}
\label{sec_Cosmological_implications} 
 
In this section, we investigate the GWs emission from the domain walls annihilation and the PBHs formation from the domain walls collapse.
 
\subsection{Gravitational waves emission} 

The GWs emitted by the domain walls annihilation can be characterized by their peak frequency and peak amplitude.
The GWs peak frequency corresponds to the Hubble parameter at the domain walls annihilation time
\begin{eqnarray}
f_{\rm peak}(t_{\rm ann}) = H(t_{\rm ann})\, ,
\end{eqnarray} 
where $f=\kappa/(2\pi) R(t)$ is the frequency with the comoving wavenumber $\kappa$, and $R(t)$ is the scale factor of the Universe.
Considering the redshift of the peak frequency due to the subsequent cosmic expansion, we can obtain the peak frequency at the present time $t_0$ as
\begin{eqnarray}
f_{\rm peak,0}\simeq 1.5\times10^{-10} \, {\rm Hz} \left(\dfrac{\zeta}{0.1}\right)^{-1/2} \left(\dfrac{\eta}{0.1}\right)^{-1} \left(\dfrac{f_a}{10^{16}\, \rm GeV}\right)^{-1/2} \left(\dfrac{\Lambda_b}{1\, \rm MeV}\right)^{2} \, .
\label{f_peak0}
\end{eqnarray} 
The GWs spectrum at the cosmic time $t$ can be described by
\begin{eqnarray}
\Omega_{\rm gw}(t,f)=\frac{1}{\rho_c(t)} \frac{{\rm d}\rho_{\rm gw}(t)}{{\rm d} \ln f}\, ,
\end{eqnarray} 
where $\rho_c(t)$ is the critical energy density.
Then the GWs peak amplitude at the domain walls annihilation time is given by \cite{Saikawa:2017hiv}
\begin{eqnarray}
\Omega_{\rm gw}(t_{\rm ann})_{\rm peak}=\frac{8\pi  \tilde{\epsilon}_{\rm gw} G^2 \mathcal{A}^2 \sigma_{\rm wall}^2}{3H^2(t_{\rm ann})}\, ,
\end{eqnarray} 
where $\tilde{\epsilon}_{\rm gw}\simeq 0.7 \pm 0.4$ is an efficiency parameter, and $G$ is the Newton’s gravitational constant.
We also have the peak amplitude at present
\begin{eqnarray}
\begin{aligned}
\Omega_{\rm gw}(t_0)_{\rm peak}h^2&=\Omega_{\rm rad} h^2 \left(\dfrac{g_{*s0}^{4/3}/g_{*0}}{g_{*\rm ann}^{1/3}}\right) \Omega_{\rm gw}(t_{\rm ann})_{\rm peak}~~\\
&\simeq 5.3\times 10^{-9} \left(\dfrac{\zeta}{0.1}\right)^4 \left(\dfrac{\eta}{0.1}\right)^8 \left(\dfrac{f_a}{10^{16}\, \rm GeV}\right)^{4} \left(\dfrac{\Lambda_b}{1\, \rm MeV}\right)^{-8}\, ,
\label{omega_peak0}
\end{aligned}
\end{eqnarray} 
where $\Omega_{\rm rad} h^2 \simeq 4.15\times 10^{-5}$ is the density parameter of radiation at present, $h\simeq 0.68$ is the reduced Hubble parameter, and $g_{*s}$ is the number of effective degrees of freedom of the entropy density.
Using eqs.~(\ref{f_peak0}) and (\ref{omega_peak0}), the present GWs spectrum is finally given by
\begin{eqnarray}
\Omega_{\rm gw}h^2=
\begin{cases}
\Omega_{\rm gw}(t_0)_{\rm peak}h^2\left(\dfrac{f}{f_{\rm peak,0}}\right)^3\, , &f \le f_{\rm peak,0}\\
\Omega_{\rm gw}(t_0)_{\rm peak}h^2\left(\dfrac{f_{\rm peak,0}}{f}\right)\, , &f > f_{\rm peak,0}~
\end{cases} 
\end{eqnarray} 
which can be described by a piecewise function that evolves as $\Omega_{\rm gw}h^2\propto f^3$ for the low frequencies, and $\Omega_{\rm gw}h^2\propto f^{-1}$ for the high frequencies \cite{Hiramatsu:2013qaa}.

\begin{figure*}[t]
\centering
\includegraphics[width=0.48\textwidth]{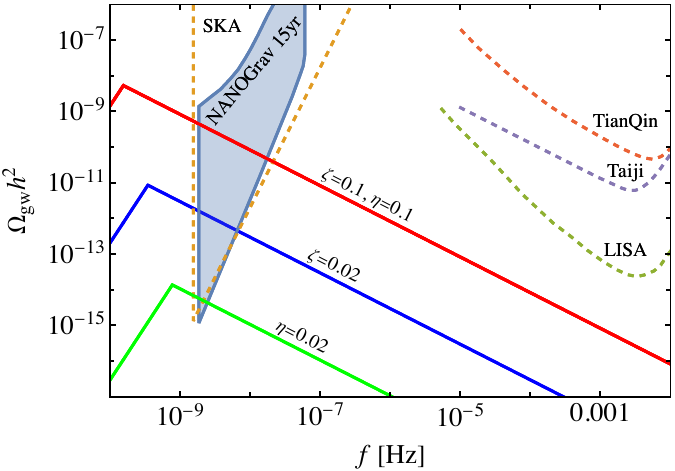}\quad\includegraphics[width=0.48\textwidth]{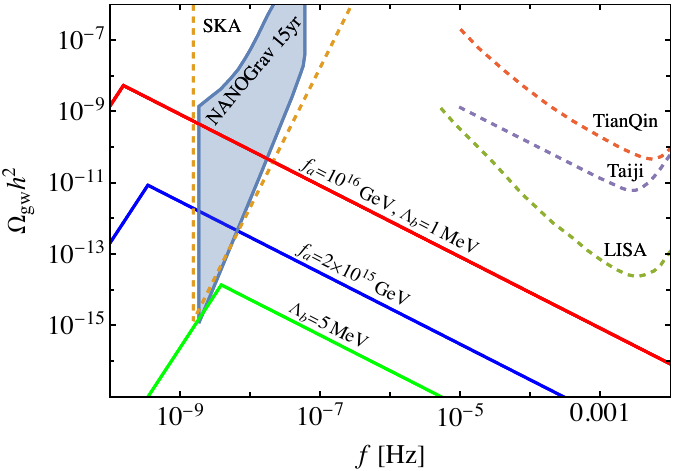}
\caption{The predicted GWs spectra $\Omega_{\rm gw}h^2$ as a function of the frequency $f$, with the parameter uncertainties of $\zeta$, $\eta$, $f_a$, and $\Lambda_b$.
The red solid line in the panels represent the result with the benchmark values $\zeta=0.1$, $\eta=0.1$, $f_a=10^{16}\, \rm GeV$, and $\Lambda_b=1\, \rm MeV$.
Left: The blue and green solid lines represent the results with $\zeta=0.02$ and $\eta=0.02$, respectively.
Right: The blue and green solid lines represent the results with $f_a=2\times10^{15}\, \rm GeV$ and $\Lambda_b=5\, \rm MeV$, respectively.
The result of the NANOGrav 15-year dataset \cite{NANOGrav:2023gor, NANOGrav:2023hfp} and other experimental sensitivities (SKA \cite{Carilli:2004nx}, TianQin \cite{TianQin:2015yph}, Taiji \cite{Ruan:2018tsw}, and LISA \cite{LISA:2017pwj}) in this plot region are also shown.}
\label{fig_gw}
\end{figure*} 

The predicted GWs spectra emitted by the axion domain walls annihilation in our scenario are shown in figure~\ref{fig_gw}.
There are four parameters ($\zeta$, $\eta$, $f_a$, and $\Lambda_b$) that significantly determine the GWs peak frequency and peak amplitude.
As discussed before, we take these benchmark values as $\zeta=0.1$, $\eta=0.1$, $f_a=10^{16}\, \rm GeV$, and $\Lambda_b=1\, \rm MeV$, corresponding to the red solid line in the panels.
While the blue and green solid lines are produced by changing only one parameter each time with $\zeta=0.02$, $\eta=0.02$, $f_a=2\times10^{15}\, \rm GeV$, and $\Lambda_b=5\, \rm MeV$, respectively.
We find that $\eta$ and $\Lambda_b$ have the significantly greater impact on the GWs spectra than $\zeta$ and $f_a$.
Note that the lines corresponding to the changes of the parameters $\zeta$ and $f_a$ have a same distribution, this is because they are degenerate in the calculations of the domain walls annihilation temperature and the GWs spectrum.
The result of the recently published NANOGrav 15-year dataset \cite{NANOGrav:2023gor, NANOGrav:2023hfp} and other experimental sensitivities (SKA \cite{Carilli:2004nx}, TianQin \cite{TianQin:2015yph}, Taiji \cite{Ruan:2018tsw}, and LISA \cite{LISA:2017pwj}) in the plot region are also shown for comparisons.
We find that the predicted nano-Hertz GWs spectra in this scenario can be tested by the current and future PTA projects.

On the other hand, according to the BBN bound and the NANOGrav 15-year dataset, we find a roughly allowed region for the bias parameter, ${0.3\, \rm MeV}\lesssim\Lambda_b\lesssim{5\, \rm MeV}$.
Note that this is obtained when other parameters ($\zeta$, $\eta$, and $f_a$) are taken as the corresponding benchmark values.
See also figure~\ref{fig_lambda_b} with the blue shadow region in the $\{f_a, \, \Lambda_b\}$ plane.
The blue line is estimated using the two points where $f_a$ equals $10^{12}\, \rm GeV$ and $10^{16}\, \rm GeV$, at which the predicted GWs spectra barely reach the NANOGrav 15-year dataset. 
Then, the shadow region, within the parameter space characterized by a larger $\Lambda_b$, indicates that the corresponding peak frequency and amplitude of GWs cannot be used to explain the observational data.
Moreover, due to the large parameter degrees of freedom in the model, and there may have the constraints on $\zeta$ and $\eta$ from the conditions for the level crossing to occur \cite{Li:2023uvt}, we do not show the concrete QCD axion (or ALP) properties ($m_a$, $f_a$) from the GWs measurement in this work.
 
\subsection{Primordial black holes formation} 
 
In this subsection, we investigate the PBHs formation from the domain walls collapse. 
During the annihilation, the closed domain walls could shrink to the Schwarzschild radius and collapse into the PBHs \cite{Ferrer:2018uiu}.
Here we consider the domain walls collapse in an approximately spherically symmetric way \cite{Gelmini:2022nim, Gelmini:2023ngs}.

\begin{figure}[t]
\centering
\includegraphics[width=0.63\textwidth]{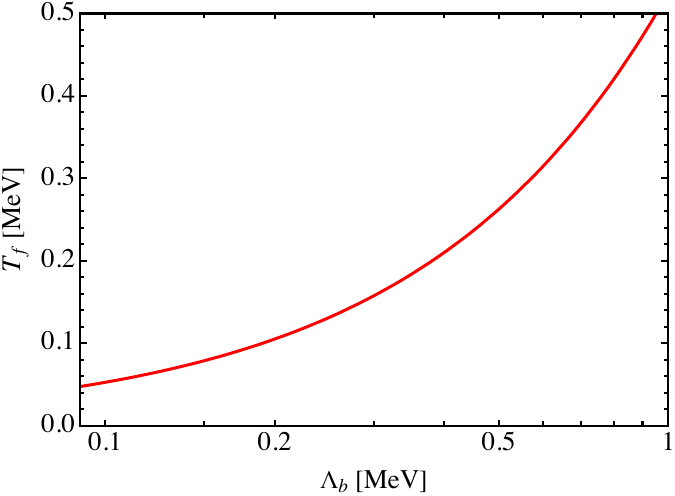}
\caption{The PBH formation temperature $T_f$ as a function of the bias parameter $\Lambda_b$.}
\label{fig_Tf}
\end{figure}

The Schwarzschild radius $R_S(t)$ at the cosmic time $t$ is given by $R_S(t)=2M(t)/M_{\rm Pl}^2$, where $M_{\rm Pl}=1.22\times10^{19}\, \rm GeV$ is the Planck mass, and $M(t)$ is the mass of the closed domain walls at the time $t$
\begin{eqnarray}
M(t)\simeq \dfrac{4}{3}\pi V_{\rm bias} t^3 + 4\pi \sigma_{\rm wall} t^2\, .
\end{eqnarray}  
The condition for PBHs to form is that the ratio of $R_S(t)$ to $t$ is close to 1, $\rm i.e.$,
\begin{eqnarray}
p(t)=\dfrac{R_S(t)}{t}=\dfrac{2M(t)}{t M_{\rm Pl}^2}\sim 1\, .
\end{eqnarray} 
At the domain walls annihilation time $t_{\rm ann}$, the walls tension pressure is comparable to the volume pressure.
Since $V_{\rm bias}\simeq \mathcal{A} \sigma_{\rm wall}/t_{\rm ann}$ that given by $p_T\simeq p_V$, we can obtain the mass $M(t)$ and the ratio $p(T)$ at $t_{\rm ann}$, $M(t_{\rm ann})\simeq 16/3 \pi V_{\rm bias} t_{\rm ann}^3$, and $p(T_{\rm ann})\simeq 30 V_{\rm bias} / (\pi^2 g_*(T_{\rm ann}) T_{\rm ann}^4)$, respectively. 
After the time $t_{\rm ann}$, since the tension pressure decreases with the time while the volume pressure remains constant with the time, the volume contribution to the density will rapidly become dominated. 
Then the mass $M(t)$ and the ratio $p(T)$ can be given by $M(t)\simeq 4/3 \pi V_{\rm bias} t^3 (1+3 t_{\rm ann}/t)$, and $p(T)\simeq p(T_{\rm ann})/4 (t/t_{\rm ann})^2 (1+3 t_{\rm ann}/t)$, respectively. 
Here the term $t_{\rm ann}/t$ can be neglected for $t\gg t_{\rm ann}$.
As mentioned before, the PBHs formation will happen when the ratio $p(T_f)\sim 1$, we have
\begin{eqnarray}
p(T_f)\simeq\dfrac{p(T_{\rm ann})}{4} \dfrac{g_*(T_{\rm ann})}{g_*(T_f)} \left(\dfrac{T_{\rm ann}}{T_f}\right)^4\sim 1\, ,
\end{eqnarray} 
where $T_f$ is the PBH formation temperature.
Then the corresponding PBH mass is given by
\begin{eqnarray}
M_{\rm PBH}\simeq \dfrac{4}{3}\pi V_{\rm bias} t_f^3\simeq  \sqrt{\dfrac{3}{32\pi V_{\rm bias}}} M_{\rm Pl}^3\, ,
\end{eqnarray} 
and the temperature $T_f$ can also be characterized as
\begin{eqnarray}
T_f\simeq\sqrt[4]{\dfrac{15 V_{\rm bias}}{2 \pi^2 g_*(T_f)}}\simeq\sqrt[4]{\dfrac{45 M_{\rm Pl}^6}{64 \pi^3 g_*(T_f) M_{\rm PBH}^2}}\, .
\end{eqnarray} 
We show the temperature $T_f$ as a function of the bias parameter $\Lambda_b$ in figure~\ref{fig_Tf}.
For the benchmark value $\Lambda_b=1\, \rm MeV$, we have $T_f\simeq 0.5\, \rm MeV$.
Note that $T_f$ is only determined by the term $V_{\rm bias}$ ($\Lambda_b$), or the PBH mass $M_{\rm PBH}$.

\begin{figure*}[t]
\centering
\includegraphics[width=0.48\textwidth]{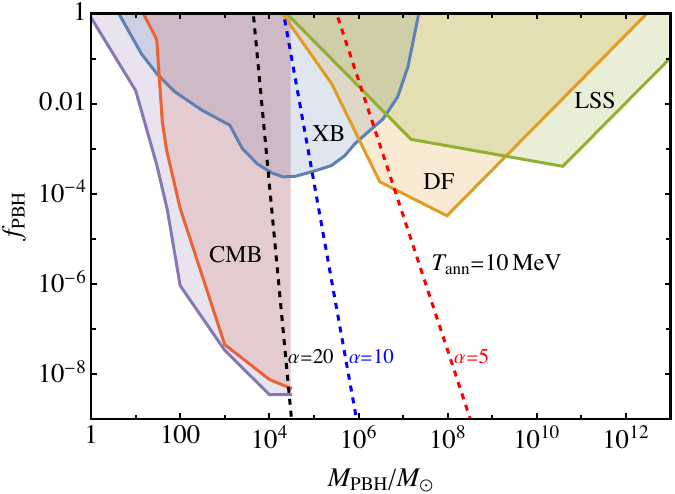}\quad\includegraphics[width=0.48\textwidth]{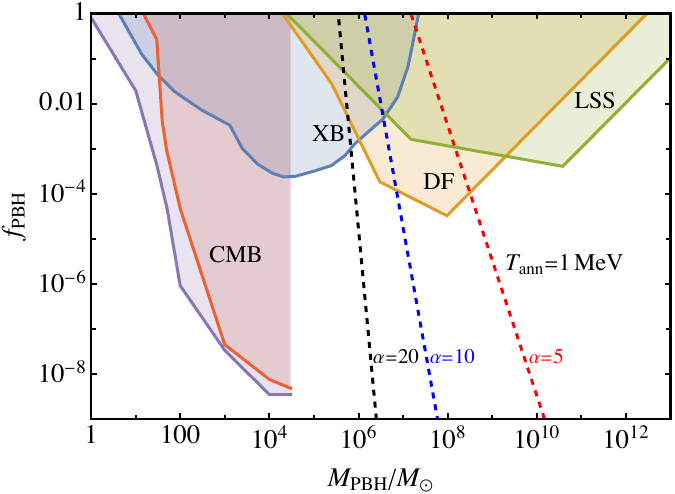}
\caption{The estimated PBH fraction $f_{\rm PBH}$ as a function of the PBH mass $M_{\rm PBH}$ (in the solar mass $M_\odot$).
Left: we set $T_{\rm ann}=10\, \rm MeV$.
Right: we set $T_{\rm ann}=1\, \rm MeV$.
The red, blue, and black dashed lines in the panels represent the PBH fraction with the typical values $\alpha=5$, 10, and 20, respectively.
The other limits (the shadow regions) are taken from the accretion limits from the CMB anisotropies measured by Planck (CMB, with two bounds) \cite{Serpico:2020ehh}, the accretion limits from the X-ray binaries (XB) \cite{Inoue:2017csr}, the dynamical limits from the infalling of halo objects due to the dynamical friction (DF) \cite{Carr:2020gox}, and the large-scale structure (LSS) limits from the various cosmic structures \cite{Carr:2018rid}.}
\label{fig_fpbh}
\end{figure*}

Then we briefly estimate the PBH fraction of the total DM energy density in this scenario.
The PBH energy density at the formation temperature is given by $\rho_{\rm PBH}(T_f)\simeq p^\beta(T_f) \rho_{\rm wall}(T_f)$, where $\beta$ is a positive factor represents the small deviations from the spherically symmetric collapse.
Since we have $p(T_f)\sim 1 \Rightarrow p^\beta(T_f)\sim 1$ and the domain walls energy density after annihilation evolves as $\rho_{\rm wall}(T)/\rho_{\rm wall}(T_{\rm ann})\simeq (T/T_{\rm ann})^\alpha$, the PBH fraction can be described by $f_{\rm PBH}\simeq (T_f/T_{\rm ann})^\alpha \rho_{\rm wall}(T_{\rm ann})/\rho_{\rm DM}(T_f)$, where $\alpha$ is a positive parameter between approximately 5 and 20 \cite{Kawasaki:2014sqa}. 
Finally, the PBH fraction is given by
\begin{eqnarray}
f_{\rm PBH}\simeq 4\left(\dfrac{45 M_{\rm Pl}^6}{64 \pi^3}\right)^{(\alpha+1)/4} \dfrac{g_{*s}(T_0)}{g_*(T_0)}\dfrac{g_*(T_f)^{(3-\alpha)/4}}{g_{*s}(T_f)}\dfrac{M_{\rm PBH}^{-(\alpha+1)/2}}{T_{\rm ann}^\alpha T_0} \dfrac{\rho_R(T_0)}{\rho_{\rm DM}(T_0)}
 \, ,
\end{eqnarray} 
where $T_0$ is the present  cosmic microwave background (CMB) temperature, $\rho_R$ is the radiation energy density, and $\rho_{\rm DM}$ is the DM energy density.
We show the estimated PBH fraction $f_{\rm PBH}$ as a function of the PBH mass $M_{\rm PBH}$ in figure~\ref{fig_fpbh}.
The other limits in this plot region from the CMB \cite{Serpico:2020ehh}, the X-ray binaries (XB) \cite{Inoue:2017csr}, the dynamical friction (DF) \cite{Carr:2020gox}, and the large-scale structure (LSS) \cite{Carr:2018rid} are also shown.
Since in our model there are four parameters ($\zeta$, $\eta$, $\Lambda_b$, and $f_a$) that determine the domain walls annihilation temperature $T_{\rm ann}$, here we just show the PBH fraction with the parameters $T_{\rm ann}$ and $\alpha$ in the figure.
We take these typical values as $T_{\rm ann}=10\, \rm MeV$ and $1\, \rm MeV$, corresponding to the left and right panels, respectively, and take $\alpha=5$, 10, and 20, corresponding to the three red, blue, and black dashed lines.
Note that we just take several typical values of the parameter $\alpha$, more accurate discussions about $\alpha$ from the simulations are required \cite{Kawasaki:2014sqa}.
We find that the PBHs in the mass range of $10^5 \, M_\odot \lesssim M_{\rm PBH} \lesssim 10^8 \, M_\odot$ could potentially form in our scenario and account for a small fraction $f_{\rm PBH}\sim 10^{-5}$ of the cold DM.

Furthermore, we also find that these PBHs may account for the seeds of SMBHs at the high redshift.
The SMBHs with the mass range of $\mathcal{O}(10^6-10^9)\, M_\odot$ are commonly found in the center of galaxies \cite{Richstone:1998ky, Mortlock:2011va, Banados:2017unc}.
However, the origin of them is not yet clear.
One may consider their formation from the stellar black holes through accretion and mergers, but it is difficult to account for the SMBHs at the high redshift $z\sim 7$.
Another scenario is considering the PBHs as their primordial origin \cite{Salpeter:1964kb, Duechting:2004dk}.
Due to the efficient accretion of matter on the massive seeds and mergings, those $\mathcal{O}(10^4-10^5)\, M_\odot$ PBHs could subsequently grow up to $\mathcal{O}(10^9)\, M_\odot$ SMBHs.
Therefore, the massive PBHs produced in our scenario are natural candidates for the seeds of the SMBHs.

\section{Conclusion}
\label{sec_conclusion}

In summary, in this work we have investigated the GWs emission and the PBHs formation from the light QCD axion scenario.
We first introduce the light QCD axion and the resulting axion level crossing, then we discuss the domain walls formation from the level crossing and their annihilation. 
Finally, we investigate the cosmological implications, including the nano-Hertz GWs emission from the domain walls annihilation and the massive PBHs formation from the domain walls collapse.
 
We consider the axion domain walls formation from the level crossing induced by the mass mixing between the light $Z_{\mathcal N}$ QCD axion and ALP, leading to the walls formation before the QCD phase transition.
A more general mixing case is considered that the heavy and light mass eigenvalues do not necessarily have to coincide with the axion masses, and there is a hierarchy between the two axion decay constants.
The conditions for domain walls formation are that, the axions should start to oscillate slightly before the level crossing, and the initial axion oscillation energy density should be large to climb over the barrier of potential.
To avoid the cosmological catastrophe, the domain walls must annihilate before dominating the Universe.
Then we focus our attention on the GWs emission and the PBHs formation.
The GWs emitted by the domain walls annihilation are determined by their peak frequency and peak amplitude.
We present the predicted GWs spectra with the peak frequency $f_{\rm peak}\sim 0.2\, \rm nHz$ and the peak amplitude $\Omega_{\rm GW}h^2\sim 5\times 10^{-9}$, which can be tested by current and future PTA projects.
During the domain walls annihilation, the closed walls could shrink to the Schwarzschild radius and collapse into the PBHs when this radius is comparable to the cosmic time.
Finally, we show the estimated PBH fraction and we find that the PBHs in the mass range of $10^5 \, M_\odot \lesssim M_{\rm PBH} \lesssim 10^8 \, M_\odot$ could potentially form in this scenario and account for a small fraction $f_{\rm PBH}\sim 10^{-5}$ of the cold DM.
Furthermore, it is natural to consider the SMBHs formation from these PBHs.

\section*{Acknowledgments}
This work was supported by the CAS Project for Young Scientists in Basic Research YSBR-006, the National Key R\&D Program of China (Grant No.~2017YFA0402204), and the National Natural Science Foundation of China (NSFC) (Grants No.~11821505, No.~11825506, and No.~12047503).

\bibliography{references}

\end{document}